\documentclass[aps,prl,twocolumn,groupedaddress,showpacs,floatfix]{revtex4}

\usepackage{graphicx}

\newcommand{\re}{{\bf r}  }

\newcommand{\nup}{n_\uparrow  }
\newcommand{\ndo}{n_\downarrow  }

\begin{document}

\title{
Electrical response of molecular systems: 
the power of self-interaction corrected Kohn-Sham theory}

\author{T. K\"orzd\"orfer, M. Mundt, and S. K\"ummel}
\affiliation{Physics Institute, University of Bayreuth, D-95440 Bayreuth, Germany}

\date{July 17, 2007}

\begin{abstract}

The accurate prediction of electronic response properties of extended molecular
systems has been a challenge for conventional, explicit density
functionals. We demonstrate that a self-interaction correction 
implemented rigorously within Kohn-Sham theory via the Optimized Effective
Potential (OEP) yields polarizabilities 
close to the ones from highly accurate
wavefunction-based calculations and exceeding the quality of
exact-exchange-OEP. The orbital structure obtained with the OEP-SIC functional
and approximations to it are discussed.

\end{abstract}


\maketitle

Gaining microscopic insight into the quantum-mechanical electronic effects
that govern energy- and charge transfer in processes like
light-harvesting, charge-separation in organic solar cells, or the response of
molecular opto-electronic devices would be extremely beneficial to the
understanding of these phenomena. But the computational complexity of solving
the many-electron Schr\"odinger equation leaves little hope that wave-function
based approaches can address these problems any time soon. The formulation of
quantum mechanics without wavefunction, i.e., density functional theory (DFT)
in the Kohn-Sham framework, is computationally much more efficient and allows
to handle systems with up to several hundreds of electrons.
Therefore, it appears as the ideal tool for
investigating the above mentioned problems. However, the predictive power of
DFT calculations depends crucially on the approximations made
in the description of the exchange-correlation effects. Structural,
ground-state molecular properties are obtained with reasonable to excellent
accuracy using standard, explicit density functionals like the Local Spin Density
Approximation (LSDA) or Generalized Gradient Approximations (GGAs). But these functionals
notoriously fail in the description of charge transfer processes
\cite{tozerCT,maitraCT} 
and associated problems like predicting the response \cite{gisbergenprl} or
transport \cite{sictransport}
properties of extended molecular systems. There is, thus,
a serious need for exchange-correlation approximations that allow to calculate
response properties like polarizabilities of extended systems reliably on a
quantitative scale and with bearable computational costs.

It has been demonstrated that improvements in the density-functional
description of the response of conjugated polymers can be achieved based on
current density functional theory \cite{faassen} and related ideas
\cite{maitrafaassen}, or by incorporating
full \cite{gisbergenprl,yangchains,kkp} or partial
\cite{sekinohirao07} exact exchange. It has also been argued
that correlation effects play a non-negligible role in the proper description
of response properties \cite{champagne0506}.
However, evaluating the Fock integrals in exact exchange approaches increases numerical costs
substantially, and the computational complexity of approaches using exact
exchange with ``compatible'' correlation is significant \cite{rmp}.

In this manuscript we demonstrate that
these problems can be overcome with a self-interaction correction (SIC)
employed rigorously within Kohn-Sham theory. In the SIC-scheme, only direct,
i.e., self-exchange integrals need to be evaluated, thus computational costs are lowered.
OEP-SIC yields highly accurate results for the response
of extended molecular systems without involving empirical parameters. 

The first ``modern'' SIC was proposed by Perdew and Zunger as a correction to
LSDA \cite{pz}. They
devised the LSDA-SIC functional
\begin{eqnarray}
E^\mathrm{SIC}_\mathrm{xc}[\nup,\ndo]&=&E^\mathrm{LSDA}_\mathrm{xc}[\nup,\ndo]-
\label{esic} \\ &&
\left[
\sum_{\sigma=\uparrow,\downarrow}\sum_{i=1}^{N_\sigma} 
E_\mathrm{H}[n_{i,\sigma}]+ E^\mathrm{LSDA}_\mathrm{xc}[n_{i,\sigma},0]
\right] \, , \nonumber
\end{eqnarray}
where $E^\mathrm{LSDA}_\mathrm{xc}$ is the LSDA exchange-correlation energy
functional, $E_\mathrm{H}$ the Hartree (classical Coulomb) energy, $\nup$ and
$\ndo$ the up- and down-spin densities, respectively, $N_\uparrow$ and
$N_\downarrow$ the numbers of occupied spin-orbitals, and $n_{i,\sigma}$ the
orbital spin densities. Eq.\ (\ref{esic}) is not the only way in which a SIC
can be defined \cite{otherSICs}, but it is plausible and 
straightforward: The spurious self-interaction effects that are contained in
the Hartree energy and the LSDA functional are subtracted on an
orbital-by-orbital basis. However, a subtlety is buried in this seemingly
simple equation: The functional depends on the orbitals explicitly, i.e., it
is not an explicit density functional. The traditional way of approaching this
problem has been to minimize the total energy with respect to the orbitals
\cite{pz,godeckersic,vydrov04}. This approach is within the realm of
the Hohenberg-Kohn theorem, but it is outside the foundations of Kohn-Sham
theory: minimizing with respect to the orbitals leads to single-particle
equations with orbital specific potentials instead of a global, local
Kohn-Sham potential for all orbitals. But the existence of a common, local
potential is one of the features that makes Kohn-Sham DFT attractive:
Only with a local potential is the non-interacting kinetic energy density a
well-defined density functional, a local potential considerably simplifies
numerical efforts, it facilitates the interpretation of results, and it yields
not only corrected occupied eigenvalues, but also corrected unoccupied ones.
But on the other hand, Perdew-Zunger SIC \cite{pz} has become popular in some
areas of solid state physics 
exactly for the reason that it does not work with one common local potential but
with several orbital specific ones, because orbital specific potentials
straightforwardly allow to take into account orbital localization
effects: SIC with orbital-specific potentials can treat, e.g., p- and 
d-orbitals of a crystalline solid on a different footing. In this way,
Perdew-Zunger SIC can naturally distinguish between localized and delocalized
states. In order to benefit from the advantages of working with a local potential without loosing
the ability to describe localization effects, schemes have been devised which
make use of the fact that Eq.\ (\ref{esic}) is not invariant under 
transformations of the orbitals that change the individual orbital densities
but leave the total density unchanged. Calculating orbitals from a common
Hamiltonian and then subject these orbitals 
to localizing transformations has proved to be a successful scheme
for solids \cite{svanesic,temmerman} and molecules
\cite{pedersonsic,garza,zieglersic}.

However, localizing orbital transformations can become computationally
involved in large finite systems and in time-dependent
calculations. Therefore, yet another variant of the SIC has become popular. It
uses the Krieger-Li-Iafrate (KLI) construction \cite{kli} to obtain the
KLI-potential corresponding to Eq.\ (\ref{esic}) and evaluates Eq.\
(\ref{esic}) directly with the KLI orbitals
\cite{kriegersic,ullrich00,graborev,chu05}. This approach has been justified
as an approximation to the OEP-version of SIC (OEP-SIC), which is defined by
evaluating Eq.\ (\ref{esic}) with the orbitals obtained from the SIC Kohn-Sham
potential that follows from the Optimized Effective Potential (OEP)
formalism \cite{rmp}. But to the best of our knowledge, the performance of the OEP-SIC
approach itself has remained largely unexplored, and tests of the KLI-SIC
approach were restricted to spherical atoms \cite{kriegersic}.
In this manuscript we demonstrate that OEP-SIC, but not KLI-SIC, allows to predict electric
response coefficients of molecular systems very reliably and may thus become
an important tool to investigate charge-transfer questions.

One of the most demanding tests of a functional's ability to adequately
describe charge transfer is calculating the polarizability of hydrogen
chains. It has been shown that obtaining the response of hydrogen chains
correctly is even more difficult than obtaining the response of real polymers like
polyacetylene \cite{faassen}. 
Therefore, calculating the polarizability of hydrogen chains has become a benchmark
test for many-particle approaches from both the density functional
\cite{gisbergenprl,faassen,maitrafaassen,yangchains,kkp} and the wave-function worlds
\cite{cham98,qmc}. Since a response quantity like the polarizability determines how a
system reacts to a field that induces a density shift, calculating the polarizability
also probes the ability to correctly describe charge transfer. As a second, positive
side effect, investigating hydrogen chains also allows us to address 
the question of size-consistency of the OEP-SIC functional \cite{pz,padvq,isoorb}.

\begin{figure*}[htb]
\includegraphics[width=5.9cm]{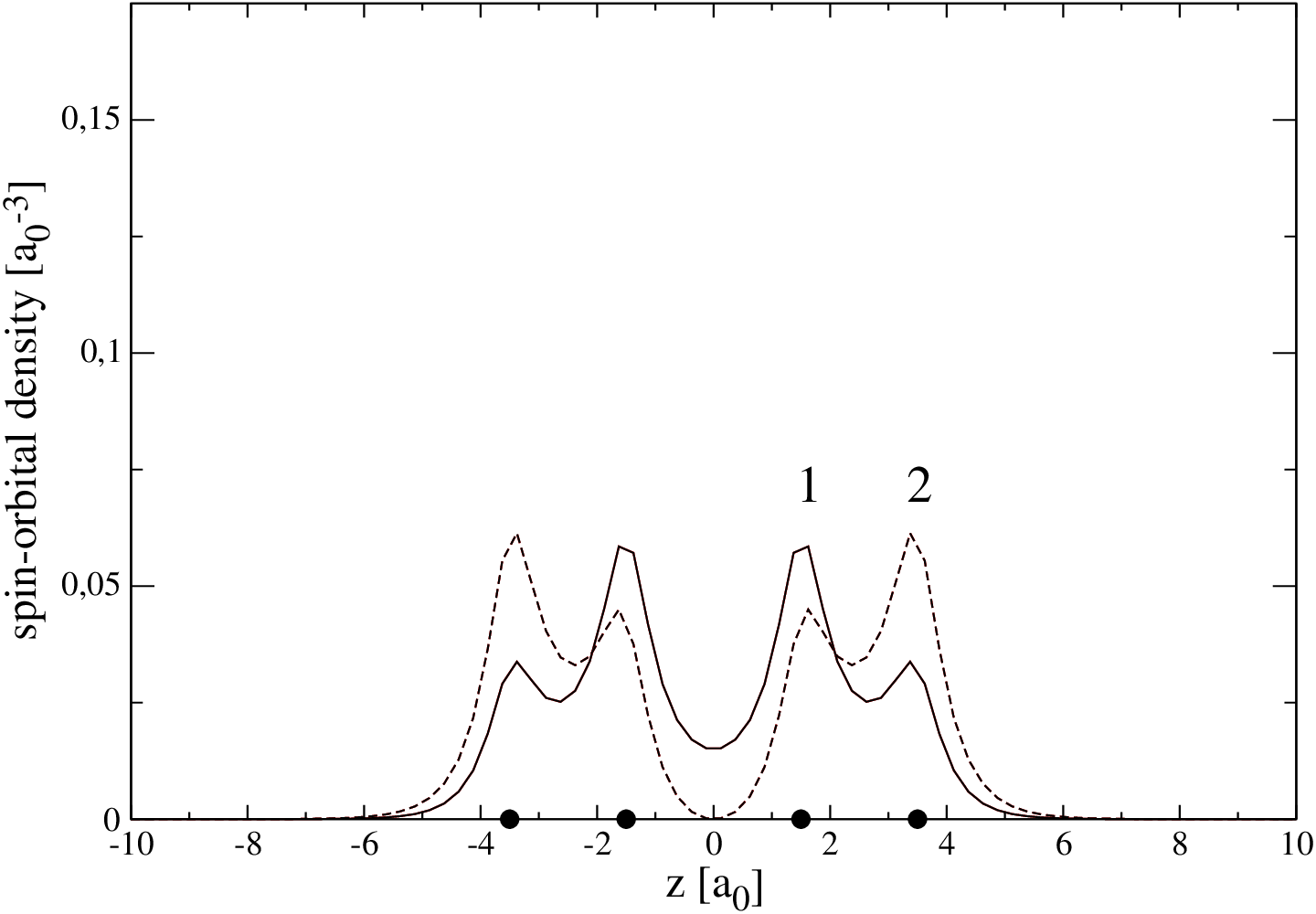}
\includegraphics[width=5.9cm]{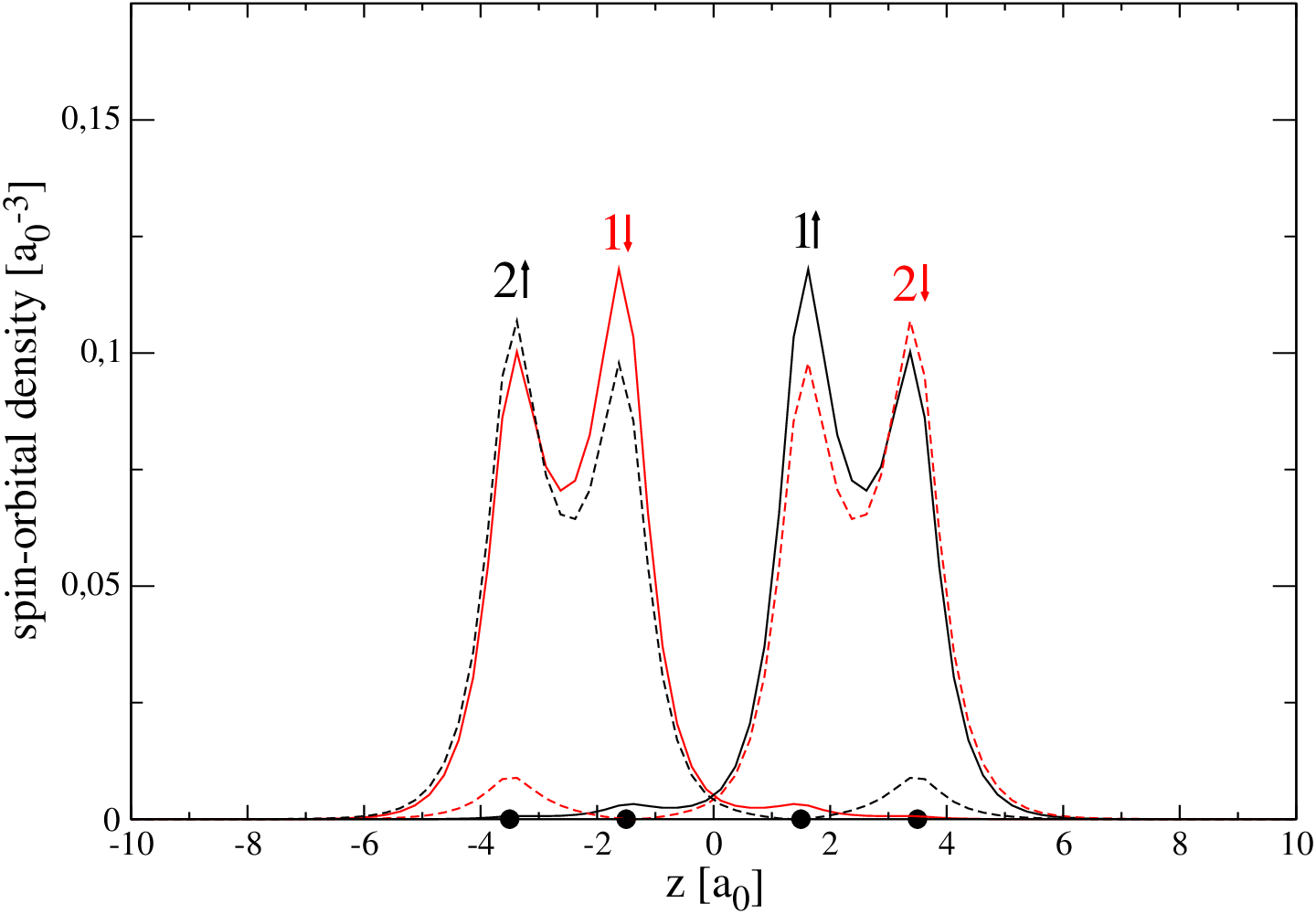}
\includegraphics[width=5.9cm]{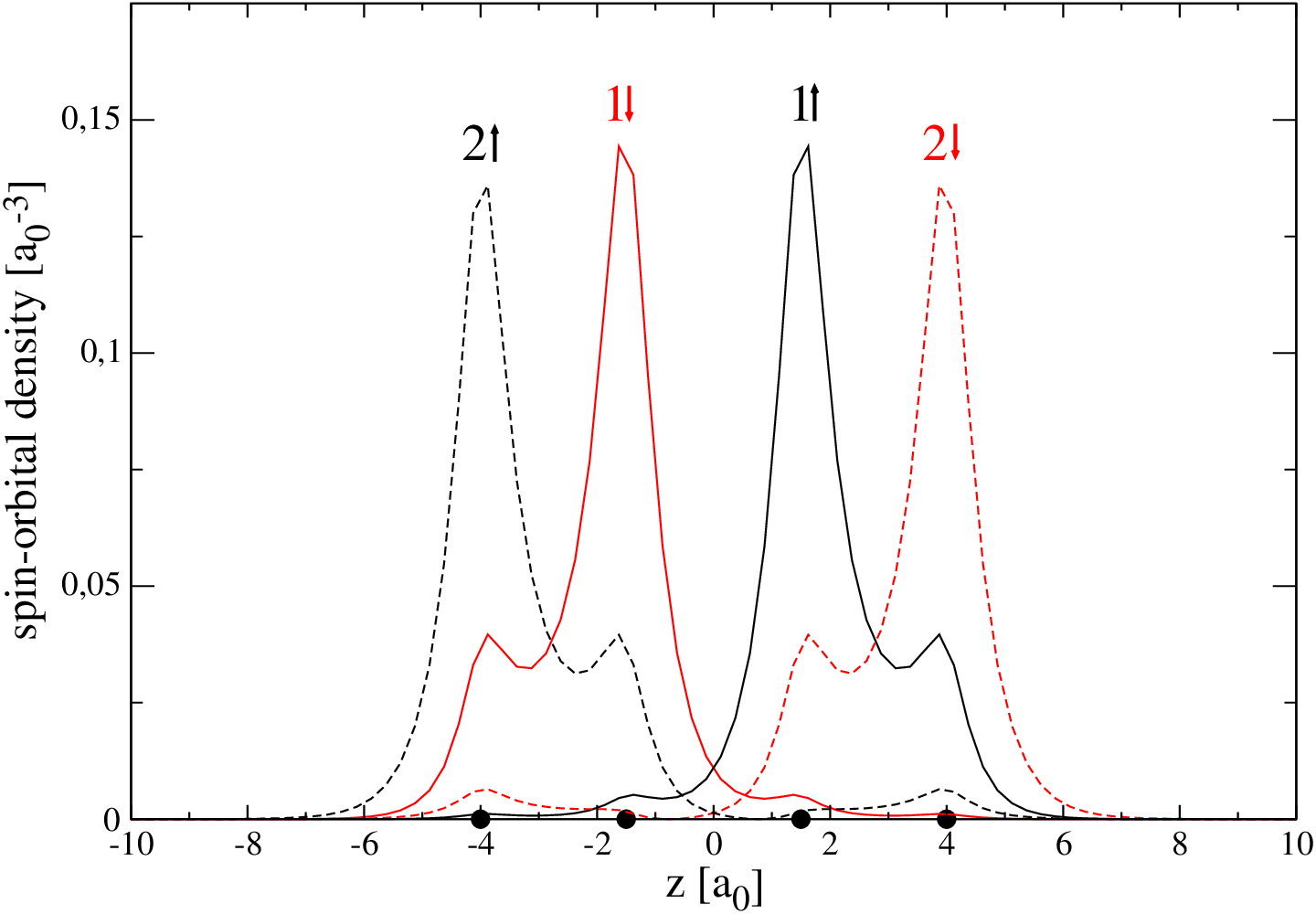}
\caption{Left: Orbital densities of $\mathrm{H}_4$ with interatomic distances
  of 2 and 3 bohr ($\mathrm{a}_0$), respectively, obtained from
  self-consistent KLI-SIC calculation. Up- and down-spin orbitals are
  identical. Middle: Spin-orbital densities for the same system obtained from
  self-consistent OEP-SIC calculation. Right: Spin-orbital densities of
  $\mathrm{H}_4$ with interatomic distances of 2.5 and 3 $a_0$, respectively, obtained from
  self-consistent OEP-SIC calculation. The orbital localization increases with
  increasing interatomic distance.
\label{fig1}}
\end{figure*}
Our calculations are based on a real space approach \cite{parsec} which we
employed to calculate the ground-state of hydrogen chains with
alternating interatomic distances of 2 and 3 $\mathrm{a}_0$ using KLI-SIC.
From the converged KLI-SIC solution we calculated the true OEP following the iterative
procedure described in \cite{itoep}, which is more cumbersone for
the SIC-LDA functional than for pure exchange, but does converge. The ground-state calculations (no
electrical field applied yet) lead to a remarkable result. For the sake of
clarity we discuss it using the specific example of the shortest chain,
$\mathrm{H}_4$. The KLI-solution is spatially symmetric as expected and as
depicted in the left part of Fig.\ \ref{fig1}. It is also manifestly
spin-unpolarized, i.e., the self-consistent KLI iteration returns to a
spin-unpolarized solution from a spin-polarized starting guess. 
But starting from the spin-unpolarized KLI solution and iterating the OEP to
self-consistency without restriction on the spin polarization, we observe a
spontaneous change in symmetry. After a few iterations of the OEP
self-consistency cycle, the up- and down-spin orbitals separate and each
orbital starts to center around one nucleus. For the interatomic
distances of 2 and 3 $a_0$ frequently used in the literature
\cite{cham98,gisbergenprl,gruening,kkp}, the effect is moderate but clearly visible, as
shown in the middle of Fig.\ (\ref{fig1}). If the interatomic distances are
increased further, e.g., to 2.5 and 3 $a_0$ as shown in the right part of Fig.\
(\ref{fig1}), the orbital localization becomes pronounced and one can
undoubtedly associate one orbital with one nucleus. This effect is not only
observed for $\mathrm{H}_4$, but also for the other hydrogen chains we
studied. 

A conclusion from this finding is
that the KLI-SIC potential is not necessarily a good approximation to the
OEP-SIC potential. In order to understand this one should recall that
the KLI-potential is justified as a mean-field approximation
\cite{kli,graborev,rmp}: The difference between the true OEP and the KLI-potential
is a term of the kind $(1/n(\re)) \nabla\cdot {\mathbf f}(\re)$, where
$\mathbf{f}(\re)$ is a well defined function depending on the full spectrum of
Kohn-Sham orbitals. Averaged over the density, the term vanishes \cite{kli,graborev}. But implicitly
this mean-field argument assumes that the ``averaged'' term has little influence in the
self-consistent iteration so that the density obtained with and without the neglected term are
very similar. However, our calculations show that this is not the case for the
SIC functional: taking into account the term that is neglected in the KLI
potential drives the self-consistent iteration to a very different solution. This is possible because the neglected term contains all orbitals and is thus relevant for unitary (in)variance and greater variational freedom. 
The breakdown of the KLI-SIC approximation may be a surprise in view of its good performace for atoms \cite{kriegersic}, but appears less surprising in view of other drawbacks \cite{zeroforce}.

The hydrogen chain ground-state results also naturally trigger
thoughts about the bulk limit that one would reach by adding ever more atoms. We briefly want to
ponder this case. Recall that for an infinite lattice of
hydrogen atoms with a lattice constant that tends to infinity, the exact Kohn-Sham
orbitals are delocalized Bloch orbitals for 
which the self-interaction correction vanishes on a per-atom basis
\cite{pz}. Using such orbitals in Eq.\ (\ref{esic}) yields the (wrong) uncorrected LSDA energy. 
Inherent to the logic of this argument is a certain order of taking the two
``infinity limits'': 
first the number of atoms tends to infinity, i.e., an infinite lattice is built, and then 
the lattice constant is made ever larger. 

Our calculations
suggest that a different result is obtained if the order of taking these two
limits is interchanged. For finite systems of largely
separated hydrogen atoms, our OEP-SIC calculations lead to localized orbitals
and thus, a self-interaction corrected energy.
Now imagine building up an ever larger lattice of hydrogen atoms with ever larger interatomic
separation by adding atoms to a finite starting system. At each step of this buildup
process, one will
be dealing with a large but finite system. Our calculations suggest
that at each
stage of the build-up process, OEP-SIC will yield localized orbitals and thus a
self-interaction corrected energy. This idea is in line with earlier findings that
revealed that it makes a great difference whether the surface of an
extended system is explicitly taken into acoount or not \cite{vanderbilt97}.
In any case our results show that OEP-SIC can yield localized orbitals that
differ greatly from the KLI orbitals. How strong the OEP-SIC localization
is will depend on the specific system. Generally speaking, we
expect localization effects to be even more pronounced in SIC schemes using
orbital dependent potentials \cite{pz} or orbital localizing transformations \cite{svanesic,temmerman,pedersonsic,garza,zieglersic}.

\begin{table}[b]
\begin{tabular}{lcccccccccc}
\hline
\hline
 &&  H$_{4}$ &&  H$_{6}$  && H$_{8}$  && H$_{10}$  && H$_{12}$ \\ 
\hline
LSDA &&  37.6  && 73 && 115 && 162  && 211 \\ 
\hline
KLI-SIC && 19.4 && 60.3 && 98.2 && 131.7 && 193.6 \\ 
\hline
OEP-EXX && 32.2  && 56.6  && 84.2  && n/a  &&138.1  \\ 
\hline
OEP-SIC && 30.6 && 48.7 && 80.1 && 98.8 && 129.8 \\ 
\hline
MP4 && 29.5 && 51.6 && 75.9 &&  n/a  && 126.9 \\
\hline
\hline
\end{tabular}
\caption{Longitudinal polarizability of hydrogen chains in $\mathrm{a}_0^3$ obtained with different
  exchange-correlation approximations. M{\o}ller-Plesset- (MP4) results taken
  from \cite{cmva}, exact-exchange OEP (OEP-EXX) from \cite{kkp}. KLI polarizabilities were
  calculated from the dipole moment, see discussion in \cite{kk}. \label{tab1}
}
\end{table}
With the ground-state structure of OEP-SIC discussed we come to the most
important aspect of this manuscript, the calculation of the electrical
response. 
As a first test we calculated the response of the two dimers $\mathrm{Na}_2$
and  $\mathrm{N}_2$, which can be seen as representing the ``extreme ends''
of dimer bonding with a soft single and a strong triple bond, respectively. 
The OEP-SIC polarizability (tensor average in $\mathrm{a}_0^3$) is obtained
as 274 for  $\mathrm{Na}_2$ (KLI-SIC performs similar) and 10.3  for $\mathrm{N}_2$ (no convergence for KLI-SIC). The value for the
sodium dimer compares favorably with the most recent experimental result of
270  \cite{rayane}, 
the value for the nitrogen dimer is too low but not unreasonable \cite{polo}. 
It is a noteworthy observation that OEP-SIC increases the polarizability (by 12\%) for
$\mathrm{Na}_2$ (where LDA underestimates) while it decreases it (by 18\%) for
$\mathrm{N}_2$ (where LDA overestimates), 
i.e., it works ``in the right direction'' in both systems. OEP-SIC also yields
greatly improved eigenvalues. 
For $\mathrm{C}\mathrm{H}_4$, e.g., OEP-SIC
yields $\varepsilon_\mathrm{HOMO}^\mathrm{OEP-SIC}=14.56$ eV, which compares
much better with the
experimental ionization energy of 14.42 eV than the LDA value
$\varepsilon^\mathrm{LDA}_\mathrm{HOMO}=9.52$ eV. 

The true and most important test is how OEP-SIC performs for the response of
extended systems where semilocal functionals fail badly. This is tested by
calculating the response of the hydrogen chains.
The Kohn-Sham SIC longitudinal static electric polarizabilities obtained from an accurate
finite field approach 
\cite{kk} are shown in Table \ref{tab1} together with LSDA, exact-exchange OEP
(OEP-EXX), and
fourth-order  M{\o}ller-Plesset perturbation theory (MP4) results. 
The MP4
results are close to the exact values and serve as the quasi-exact
benchmark. The first observation is that the KLI-SIC results vary
unsystematically -- the polarizability of $\mathrm{H}_4$ is substantially
underestimated, whereas the polarizability of all other chains is
overestimated. Comparison with OEP-EXX and LSDA shows
that KLI-SIC improves over LSDA, but is less accurate than 
exchange-only theory. The picture changes
when SIC is employed with the true, self-consistent OEP instead of
with the KLI-approximation: KLI-SIC and OEP-SIC polarizabilities are rather
different. 
Comparing OEP-SIC to the wavefunction based
results shows that OEP-SIC polarizabilities are within a few percent of the
MP4 results in all cases and are noticeably closer to the MP4 values than the
exchange-only OEP results, which up to now represented the best density functional
results for such systems. 

One may wonder why the SIC functional, in which localization of the orbitals
plays an important role, and exact exchange, which is unitarily invariant and
thus independent of orbital localization, can both lead to a reasonable
description of the chain response. The solution lies in the interpretation of
the exchange part of the SIC functional: The 
Hartree self-interaction
correction corresponds to the
self-exchange part of the EXX functional, and it is well known that
the diagonal (self-)exchange integrals are the dominant part of exchange,
i.e., they are noticeably larger than the off-diagonal
exchange integrals. 
The larger the diagonal ``classical'' parts of the
exchange energy are
in comparison with its off-diagonal parts, the more accurate becomes the SIC
description which neglects the off-diagonal parts. Since the diagonal parts
are typically maximal for localized orbitals, it becomes clear why localized orbitals are
crucial in the SIC-approach.
So from this perspective, SIC takes into account the
most important part of EXX at the cost of needing to employ localized
orbitals, but with the huge benefit of greatly reducing the number of exchange
integrals that have to be evaluated. In addition, SIC offers an improvement over bare
EXX that can be attributed to 
the non-EXX parts of the functional.
Following \cite{gisbergenprl} one can also show that the improved
OEP-SIC polarizabilities stem from  a field-counteracting-term in the 
response-part of the exchange-correlation-potential \cite{diplomtk}.
Thus, OEP-SIC is an approach which allows to reliably investigate the
electrical response of a broad range of molecular systems.  

Note added after submission: We learned of studies by Pemmaraju, Sanvito, and Burke, and
Ruzsinszky, Perdew, Csonka, Scuseria,
and Vydrov, which also find that SIC improves the response.  
\begin{acknowledgments}
S.K. acknowledges financial support by the German-Israel
Foundation.
\end{acknowledgments}

\end{document}